\documentclass[12pt]{article}

\begin{document}

\title{Negative Volterra Flows}
\author{G.M.Pritula${}^{1}$ and V.E.Vekslerchik${}^{1,2}$
\thanks{Regular Associate of the
   Abdus Salam International Centre for Theoretical Physics,
   Trieste, Italy}
}
\date{\today}
\maketitle

\begin{center}
{\it
  ${}^{1}$ Institute for Radiophysics and Electronics, Kharkov, Ukraine

  ${}^{2}$ Universidad de Castilla-La Mancha, Ciudad Real, Spain
}
\end{center}

\begin{abstract}
Taking the standard zero curvature approach we derive an infinite 
set of integrable equations, which taken together form the 
negative Volterra hierarchy. The resulting equations turn out to 
be nonlocal, which is usual for the negative flows. However, in 
some cases the nonlocality can be eliminated. Studying the 
combined action of both positive (classical) and negative Volterra 
flows, i.e. considering the differential consequences of equations 
of the extended Volterra hierarchy, we deduce local equations 
which seem to be promising from the viewpoint of applications. The 
presented results give answers to some questions related to the 
classification of integrable differential-difference equations. We 
also obtain dark solitons of the negative Volterra hierarchy using 
an elementary approach.

\end{abstract}

\section{Introduction.} \label{sec-intro}

The present paper is devoted to an integrable hierarchy which can
be viewed as an extension of the Volterra hierarchy (VH) by taking
into account the so-called 'negative' flows. The idea of negative
flows is not new. One of the first examples (and maybe the most
striking one) is the relation between the AKNS and sine-Gordon 
models \cite{N}: the sine-Gordon equation can be derived as the 
negative flow of the AKNS hierarchy. Some recent results and 
approaches to this problem can be found in \cite{DGS,AFGZ,QCS,Q}. 
The negative flows have been constructed for almost all known 
integrable systems. But there are a few exceptions, and the 
Volterra model seems to be among them: to our knowledge, the 
systematic derivation of its 'negative' equations has not been 
discussed in the literature.

Some of the equations belonging to the extended Volterra hierarchy 
are already known. First, it contains the famous 2D Toda lattice 
\cite{H}. We would also like to mention the works of De Lillo and 
Konotop \cite{dLK}, who discussed a nonlocal modification of the 
Volterra chain and recent papers by Boiti \textit{et al} 
\cite{BPPS,BBPP}, who derived and studied new integrable 
discretization of KdV. As will be shown below, these equations 
belong to the extended VH.

Another indication of the fact that the neagtive Volterra flows
should be studied more carefully is related to the problem of
classification of the integrable discrete equations. In the paper
by Yamilov \cite{Y} it has been shown that all equations of the
form

\begin{equation}
  u_{n,t} = f(u_{n+1}, \ u_n, \ u_{n-1})
\label{ide}
\end{equation}
can be divided into three classes. One of them can be presented as

\begin{equation}
  u_{n,t} =
  {p(u_n) u_{n+1} u_{n-1} +
  q(u_n) (u_{n+1} + u_{n-1}) +
  r(u_n) \over u_{n+1} - u_{n-1}}
\label{ideIII}
\end{equation}
where $p$, $q$ and $r$ are polynomials in $n$. This equation is
known to satisfy integrability tests, to posses an infinite set of
local conservation laws and generalized symmetries. However, as
was stated, for example, in \cite{LY}, "it is the only example of
nonlinear chain of the form (\ref{ide}) which cannot be reduced to
the Toda or Volterra equations by Miura transformations". In this
paper we will show that there is no need to look for substitutions 
converting (\ref{ideIII}) into Volterra chain, equations 
(\ref{ideIII}) (at least the simplest of them) are nothing but 
equations of the extended VH. 

The most straightforward approach to the negative flows is to
consider them in the framework of the inverse scattering
transform. In section \ref{sec-zcr} we start with the standard
zero-curvature representation (ZCR) for the VH and derive
equations of the negative Volterra subhierarchy. The resulting
equations turn out to be nonlocal. It is usual for the case of
negative flows. However, nonlocality does not mean that equations
are not interesting from the viewpoint of possible applications. 
Moreover, sometimes nonlocality can be eliminated. This can be 
achieved by a redefinition of the variables or by considering some 
differential consequences of the equations in question. Namely 
this is the topic of section \ref{sec-models} where we present the 
local equations which in this or another way are related to the 
nonlocal Volterra equations. Finally, in section 
\ref{sec-solitons} we derive the soliton solutions for the 
negative VH.

\section{Zero curvature representation.} \label{sec-zcr}

The inverse scattering approach for integrable systems is based on
the ZCR, when our nonlinear equations are presented as a
compatibility condition for some linear system. For example, the
Volterra chain,

\begin{equation}
  \dot{u}_{n} = u_{n} \left( u_{n+1} - u_{n-1} \right),
  \qquad
  \qquad
  n = 0, \pm 1, \pm 2, ...
\label{volterra-chain}
\end{equation}
is the compatibility condition,

\begin{equation}
  \dot{U}_{n} = V_{n+1}U_{n} - U_{n}V_{n},
\label{zcr}
\end{equation}
for the system \cite{FT}

\begin{equation}
  \left\{
  \begin{array}{lcl}
  \Psi_{n+1} &=& U_{n} \Psi_{n}
  \\
  \dot\Psi_{n} &=& V_{n} \Psi_{n}
  \end{array}
  \right.
\label{linear-system}
\end{equation}
where

\begin{equation}
U_{n} = U_{n}(\lambda) = \pmatrix{ \lambda & u_{n} \cr -1 & 0 },
\qquad
\qquad
V_{n} = V_{n}(\lambda) =
  \pmatrix{
    u_{n} & \lambda u_{n} \cr
    - \lambda & u_{n-1} - \lambda^{2} }
\label{volt-chain-uv-pair}
\end{equation}
and $\lambda$ is an arbitrary constant. Traditionally Volterra and
Toda chains are considered more often in the framework of the
'big' Lax representation, when the system is finite ($n=1,...,N$)
and $U$ and $V$ are $N \times N$ matrices. In this paper we will
use the $2 \times 2$ $U$-$V$ pair (\ref{volt-chain-uv-pair}). This 
approach is equivalent to the $N \times N$ representation and 
sometimes even more convenient (that is, it can be more easily 
modified to the cases of different boundary conditions, including 
the soliton case when $N=\infty$).

The choice of the $U$-$V$ pair for the system
(\ref{linear-system}), even for a given $U$-matrix, is not unique.
The matrix $V$ in (\ref{volt-chain-uv-pair}) is a second-order
polynomial in $\lambda$. By simple algebra one can find other
matrices $V$, which are polynomials of higher order, such that ZCR
(\ref{zcr}) will be satisfied for all $\lambda$ provided $u_{n}$ 
solve some nonlinear evolutionary equations. These equations are 
called 'higher Volterra equations'. Taken together they constitute 
the VH.

Now we are going to derive the negative Volterra flows. The idea
is simple: let us search for the $V$-matrices which are
polynomials not in $\lambda$ but in $\lambda^{-1}$. For example,
the simplest of such $V$-matrices, which is linear in
$\lambda^{-1}$, is given by

\begin{equation}
  V_{n} =
  \pmatrix{ 
    0 & -\frac{p_{n-1}}{\lambda} \cr 
    \frac{1}{\lambda p_{n-1}}  & \frac{1}{p_{n-1}} }
\end{equation}
It is easy to verify that (\ref{linear-system}) will be consistent
if $u_{n}$ and $p_{n}$ satisfy the system

\begin{equation}
  \left\{
  \begin{array}{rcl}
  \dot{u}_{n}  & = & p_{n-1} - p_{n}
  \cr
  u_{n} &=& p_{n-1} p_{n}
  \end{array}
\right.
\end{equation}
This is the first negative Volterra equation.

Now our aim is to obtain an infinite set of the similar
$V$-matrices which solve ZCR (\ref{zcr}). Using the notation

\begin{equation}
  V_{n} = \pmatrix{ a_{n} & b_{n} \cr c_{n} & d_{n} }
\end{equation}
we can rewrite (\ref{zcr}) as
\begin{eqnarray}
  0 &=&
    \lambda \left( a_{n+1} - a _{n} \right) - b_{n+1} - u_{n}c_{n},
  \\
  0 &=& b_{n} + u_{n}c_{n+1},
  \\
  0 &=& \lambda c_{n+1} + a _{n} - d_{n+1}
\end{eqnarray}
and

\begin{equation}
    \dot u_{n} = u_{n} \left( a_{n+1} - d _{n} \right) - \lambda b_{n}
\end{equation}
or, after eliminating $b_{n}$ and $d_{n}$,
\begin{eqnarray}
  0 &=&
    \lambda \left( a_{n+1} - a _{n} \right) +
    u_{n+1} c_{n+2} - u_{n}c_{n},
\label{zcr-1a}
  \\
  \dot u_{n} &=&
  u_{n} \left[ a_{n+1} - a_{n-1} + \lambda \left( c_{n+1} - c _{n} \right)\right]
\label{zcr-1b}
\end{eqnarray}
It is easy to show that (\ref{zcr-1a}) and (\ref{zcr-1b}) possess 
solutions where $a_{n}$ are polynomials of the $(2j-2)$th order in 
$\lambda^{-1}$ while $c_{n}$ are polynomials of the $(2j-1)$th 
order for $j=1,2,...$. In what follows we indicate different 
polynomials with the upper index, $a_{n}^{(j)}, c_{n}^{(j)}$, and 
introduce an infinite set of times, $\bar t_{j}$, to distinguish 
the resulting nonlinear equations.

By simple algebra one can establish the following relations
between different polynomials:
\begin{eqnarray}
  a_{n}^{(j+1)} &=&
  \lambda^{-2} a_{n}^{(j)} +
  \lambda^{-2} \alpha_{n}^{(j)},
\label{def-alpha}
  \\
  c_{n}^{(j+1)} &=&
  \lambda^{-2} c_{n}^{(j)} +
  \lambda^{-1} \gamma_{n}^{(j)}
\label{def-gamma}
\end{eqnarray}
where $\alpha_{n}^{(j)}$ and $\gamma_{n}^{(j)}$ do not depend on
$\lambda$. Substituting (\ref{def-alpha}) and (\ref{def-gamma}) 
into (\ref{zcr-1a}) and (\ref{zcr-1b}) one can convert our system 
into
\begin{eqnarray}
&&
  \alpha_{n+1}^{(j)} - \alpha_{n}^{(j)}
  + u_{n+1} \gamma_{n+2}^{(j)}
  - u_{n}   \gamma_{n}^{(j)}
  = 0,
\label{zcr-alpha}
\\&&
  \gamma_{n}^{(j)} +
  \alpha_{n}^{(j+1)} + \alpha_{n-1}^{(j+1)}
  = 0
\label{zcr-gamma}
\end{eqnarray}
and

\begin{equation}
  { \partial \over \partial \bar t_{j} } \ln u_{n} =
    \alpha_{n-1}^{(j)} - \alpha_{n+1}^{(j)}
\end{equation}
After introducing the tau-functions of the VH,

\begin{equation}
  u_{n} = { \tau_{n+1} \tau_{n-2} \over \tau_{n} \tau_{n-1} }
\label{def-tau}
\end{equation}
equations (\ref{zcr-alpha}) and (\ref{zcr-gamma}) can be solved,
\begin{eqnarray}
  \alpha_{n}^{(j)}  & = &
  { \partial \over \partial \bar t_{j} } \,
  \ln { \tau_{n-1} \over \tau_{n} }
\\
  \gamma_{n}^{(j)}  & = &
  { \partial \over \partial \bar t_{j+1} } \,
  \ln { \tau_{n} \over \tau_{n-2} }
\end{eqnarray}
and the resulting equations become

\begin{equation}
  \tau_{n-1} \tau_{n+1} \,
  { \partial \over \partial \bar t_{j+1} } \,
  \ln { \tau_{n+1} \over \tau_{n-1} } +
  \tau_{n}^{2} \,
  {\partial^{2} \over \partial\bar t_{1} \partial\bar t_{j} }
  \ln \tau_{n} = 0 
  \label{neg-hie}
\end{equation}
together with the first one

\begin{equation}
  \tau_{n-1} \tau_{n+1} \,
  { \partial \over \partial \bar t_{1} } \,
  \ln { \tau_{n+1} \over \tau_{n-1} } =
  \tau_{n}^{2}.
\label{neg-hie-1}
\end{equation}
These equations can be rewritten in the Hirota bilinear form as

\begin{equation}
  2 D_{j+1} \; \tau_{n+1} \cdot \tau_{n-1} +
  D_{j} D_{1} \; \tau_{n} \cdot \tau_{n} = 0
\label{hie-Hirota}
\end{equation}
and

\begin{equation}
  D_{1} \; \tau_{n+1} \cdot \tau_{n-1} =
  \tau_{n}^{2}
\label{hie-Hirota-1}
\end{equation}
where

\begin{equation}
D_{j} \; a \cdot b
  =
  \left.
  {\partial \over \partial \xi } \;
  a\left(\dots , \bar t_{j} + \xi , \dots \right)
  b\left(\dots , \bar t_{j} - \xi , \dots \right)
  \right|_{\xi=0}.
\end{equation}

\section{Some integrable models.} \label{sec-models}

In the previous section we have obtained an infinite set of
integrable equations. These equations are nonlocal and it is
difficult to say whether they are of any importance, say, from the 
viewpoint of applications. However, a hierarchy is more than a set 
of equations. The key point is that all equations of a hierarchy 
are compatible (some notes on this question can be found in the 
appendix). Hence we can consider them simultaneously and study the 
combined action of different flows. In other words, we can take 
some finite set of equations of a hierarchy and analyse equations 
which follow from this system. So, this is the aim of this 
section. Starting with the equations of the VH (both negative, 
derived above, and positive, i.e. classical Volterra equations) we 
will deduce some their consequences, which are local and some of 
which seem to be more promising for applications.

\subsection{2D Toda lattice and the VH.} \label{model-Toda}

First let us consider the simplest negative equation
(\ref{neg-hie-1}) together with the Volterra equation
(\ref{volterra-chain})
\begin{eqnarray}
  { \partial \over \partial \bar{t} } \; \ln { \tau_{n+1} \over \tau_{n-1} }
  & = &
  { 1 \over p_{n} }
\label{Toda-pos}
  \\
  { \partial \over \partial t } \; \ln u_{n}
  & = &
  u_{n+1} - u_{n-1}
\label{Toda-neg}
\end{eqnarray}
where $p_{n}$ is related to $\tau_{n}$ by

\begin{equation}
  p_{n} = { \tau_{n-1} \tau_{n+1} \over \tau_{n}^{2} }.
\end{equation}
One can derive from the second equation that

\begin{equation}
  { \partial \over \partial t } { 1 \over p_{n} } =
  p_{n-1} - p_{n+1}
\end{equation}
which leads to

\begin{equation}
  { \partial^{2}  \over \partial t \, \partial \bar{t} } \;
  \ln \tau_{n}  =
  - { \tau_{n-1} \tau_{n+1} \over \tau_{n}^{2} }
\label{eq-Toda}
\end{equation}
where an unessential constant has been omitted. This equation,
which can be rewritten in terms of $p_{n}$ as

\begin{equation}
  { \partial^{2}  \over \partial t \, \partial \bar{t} } \; \ln p_{n}  =
  - p_{n-1} + 2 p_{n} - p_{n+1}
\end{equation}
is nothing but the 2D Toda lattice \cite{H}. So, we have shown 
explicitly that tau-function (\ref{def-tau}) of the VH is at the 
same time the tau-function of the 2D Toda lattice, or, in other 
words, that the 2D Toda equation can be embedded into the VH. The 
reverse statement seems not to be true: not every solution of 
(\ref{eq-Toda}) should solve (\ref{Toda-pos}) and 
(\ref{Toda-neg}). In this sense splitting the (2+1)-dimensional 
equation (\ref{eq-Toda}) into two (1+1)-dimensional equations from 
the VH is a kind of \emph{ansatz}. But the class of solutions of 
the 2D Toda model which can be can be obtained by solving Volterra 
equations (\ref{Toda-pos}) and (\ref{Toda-neg}) contains a large 
number of important solutions such as, e.g., N-soliton and 
finite-gap quasiperiodic solutions. It is obvious that this is a 
relation between the 2D Toda model and the \emph{extended} VH and 
could not be revealed if dealing with the classical Volterra 
equations only.

\subsection{Singular chain.}

Another consequence of (\ref{neg-hie-1}) is that the quantity

\begin{equation}
  x_{n} = { \partial \over \partial\bar t_{1}} \ln \tau_{n}
\end{equation}
satisfies

\begin{equation}
  \dot{x}_{n} = { 1 \over x_{n-1} - x_{n+1} }
\label{sing-evol}
\end{equation}
where the dot stands for the differentiating with respect to the 
first 'positive' Volterra time (i.e. time which corresponds to 
(\ref{volterra-chain})), $t$, $\dot{x}_{n} = \partial x_{n} / 
\partial t$. Thus we have come to an equation of the type 
(\ref{ideIII}) mentioned in the introduction. So, equations 
(\ref{ideIII}) probably can be embedded in the extended VH. Note 
that the quantity $x_{n}$ cannot be expressed locally in terms of 
the 'positive' Volterra subhierarchy, without invoking the first 
negative flow.

Equation (\ref{sing-evol}) seems to be non-typical for physical
aplications. However, it can be easily converted to a more usual
form. Indeed, it follows from (\ref{sing-evol}) that the
quantities $Q_{n}$ and $P_{n}$, given by

\begin{equation}
  Q_{n} = { 1 \over 2} \ln { x_{n+1} \over x_{n-1} }
  \qquad
  P_{n} = x_{n+1} x_{n}
\end{equation}
satisfy the system
\begin{eqnarray}
2 \dot Q_{n} &=&
  { 1 \over P_{n} - P_{n+1} } + { 1 \over P_{n} - P_{n-1} }
\\
2 \dot P_{n} &=&
  - \coth\left( Q_{n} + Q_{n+1} \right)
  - \coth\left( Q_{n} + Q_{n-1} \right)
\end{eqnarray}
which is a Hamiltonian system

\begin{equation}
  \dot Q_{n} = { \partial \mathcal{H} \over \partial P_{n} }
  \qquad
  \qquad
  \dot P_{n} = - { \partial \mathcal{H} \over \partial Q_{n} }
\end{equation}
with the standard Poisson bracket and the Hamiltonian given by

\begin{equation}
\mathcal{H} = {1 \over 2} \sum_{n} \left\{
  \ln\left| \, P_{n} - P_{n-1} \, \right| +
  \ln\left| \, \sinh\left( Q_{n} + Q_{n-1} \right) \, \right|
  \right\}.
\end{equation}

\subsection{Integrable dynamics of roots of a 3-polynomial.}

The following system is a 'toy' model which can be derived from
(\ref{sing-evol}) by imposing the periodic conditions
$x_{n+3}=x_{n}$. Noting that the product

\begin{equation}
  \Pi =
  \left( x_{1} - x_{2} \right)
  \left( x_{2} - x_{3} \right)
  \left( x_{3} - x_{1} \right)
\end{equation}
is a constant and rescaling the time $t \to \Pi \cdot t$ one can 
rewrite (\ref{sing-evol}) as

\begin{equation}
  \left\{
  \begin{array}{l}
  \dot{x}_{1} = \left( x_{1} - x_{2} \right)\left( x_{1} - x_{3} \right) \cr
  \dot{x}_{2} = \left( x_{2} - x_{1} \right)\left( x_{2} - x_{3} \right) \cr
  \dot{x}_{3} = \left( x_{3} - x_{1} \right)\left( x_{3} - x_{2} \right)
  \end{array}
  \right.
\end{equation}
or

\begin{equation}
  \dot x_{k} = \mathcal{P}'(x_{k}) \qquad k=1,2,3
\end{equation}
where $\mathcal{P}$ is an arbitrary monic 3-polynomial, and
$x_{k}$ are its roots:

\begin{equation}
  \mathcal{P}(x) = \prod_{k=1}^{3} \left( x - x_{k} \right).
\end{equation}
Of course, it is a very simple ordinary differential equation,
which can be studied without using the inverse scattering
transform machinery. In terms of the first symmetric function of 
the roots, $ \sigma = x_{1} + x_{2} + x_{3} $, it can be rewritten 
as the stationary KdV,

\begin{equation}
  \sigma_{ttt} - 6\sigma_{t}^{2} = 0
\end{equation}
(to make the following formulae more readable we indicate
differentiatings with subscripts) or, after introducing a
'logarithmic potential' $\psi$,

\begin{equation}
  \psi_{t} / \psi = - \sigma,
\end{equation}
as

\begin{equation}
  \psi_{tttt} \, \psi - 4 \psi_{ttt} \, \psi_{t} + 3 \psi_{tt}^{2} = 0
\end{equation}
which is the expanded form of the simple bilinear equation

\begin{equation}
  D_{t}^{4} \; \psi \cdot \psi= 0.
\end{equation}

\subsection{3+1 dimensional example.}

The last example we want to discuss differs from those presented 
above. All previous equations considered in this sectioin were 
difference-differential systems. Now we derive from the VH some 
partial derivative equations in (3+1)-dimensional (physical) 
space-time.

Out starting point is equation (\ref{neg-hie}) which can be
presented as

\begin{equation}
  p_{n} \;
  {\partial \over \partial\bar{t}_{j+1} }  \;
  \ln { \tau_{n+1} \over \tau_{n-1} } =
  -  {\partial^{2}  \over \partial\bar{t}_{j}\partial\bar{t}_{1} } \;
  \ln \tau_{n}.
\label{neg-hie-p}
\end{equation}
Acting on both sides of this equation by the $ \partial /
\partial\bar{t}_{k} $ operator one can note that right-hand side
is symmetrical in $j$ and $k$, which leads to

\begin{equation}
  \left(
    {\partial \over \partial\bar{t}_{j} }  \;
    p_{n} \;
    {\partial \over \partial\bar{t}_{k+1} }
  -
    {\partial \over \partial\bar{t}_{k} }  \;
    p_{n} \;
    {\partial \over \partial\bar{t}_{j+1} }
  \right)
  \ln { \tau_{n+1} \over \tau_{n-1} } = 0.
\label{eq-asym}
\end{equation}
On the other hand, differentiating (\ref{neg-hie-p}) with respect 
to the first 'positive' Volterra time $t$ one gets

\begin{equation}
  {\partial \over \partial t }  \;
  p_{n} \;
  {\partial \over \partial\bar{t}_{j+1} }  \;
  \ln { \tau_{n+1} \over \tau_{n-1} } =
  -  {\partial  \over \partial\bar{t}_{j} } \;
  \left( {\partial^{2}  \over \partial t \, \partial\bar{t}_{1} } \;
  \ln \tau_{n} \right)
\end{equation}
which can be rewritten using (\ref{neg-hie-1}) as

\begin{equation}
  {\partial \over \partial t }  \;
  p_{n} \;
  {\partial \over \partial\bar{t}_{j+1} }  \;
  \ln { \tau_{n+1} \over \tau_{n-1} } =
  {\partial  \over \partial\bar{t}_{j} } \; p_{n}.
\label{eq-time}
\end{equation}
It is easy to note that equations (\ref{eq-asym}) and
(\ref{eq-time}) form a closed system
\begin{eqnarray}
&&
  {\partial \over \partial\bar{t}_{j} }  \; p \;
  {\partial \Lambda \over \partial\bar{t}_{k+1} }
  -
  {\partial \over \partial\bar{t}_{k} }  \; p \;
  {\partial \Lambda \over \partial\bar{t}_{j+1} }
  = 0
\label{eq-asym-n}
\\[3mm]
&&
  {\partial \over \partial t }  \; p \;
  {\partial \Lambda \over \partial\bar{t}_{j+1} }  =
  {\partial  p \over \partial\bar{t}_{j} }
\label{eq-time-n}
\end{eqnarray}
for two functions

\begin{equation}
  p = p_{n}
  \;\;
  \mathrm{and}
  \;\;
  \Lambda = \ln { \tau_{n+1} \over \tau_{n-1} }
  \qquad
  n = \mathrm{constant}
\end{equation}
with $n$ being fixed.

From this multidimensional system one can derive a rather
interesting consequence in 3+1 dimensions. To this end we denote

\begin{equation}
  \left(x, \; y, \; z \right)
  =
  \left( \bar t_{1}, \; \bar t_{2}, \; \bar t_{3} \right)
\end{equation}
and introduce vector $\vec{u}$ by

\begin{equation}
  \vec{u} =
  \pmatrix{
  \displaystyle
    {\partial \Lambda \over \partial\bar{t}_{2}}, \;
    {\partial \Lambda \over \partial\bar{t}_{3}}, \;
    {\partial \Lambda \over \partial\bar{t}_{4}}
  }^{T}.
\end{equation}
In these terms equation (\ref{eq-time-n}) can be rewritten as

\begin{equation}
  {\partial \over \partial t }  \; p \, \vec{u} =
  \nabla p
\label{eq-nabla}
\end{equation}
where $\nabla$ is the three-dimensional gradient,
  $\nabla = \left(
  \partial/\partial x, \partial/\partial y,\partial/\partial z
  \right)$,
while equation (\ref{eq-asym-n}) gives

\begin{equation}
  \mathrm{curl} \; p \, \vec{u} = 0.
\label{eq-curl}
\end{equation}
Eliminating $p$ from (\ref{eq-nabla}) and (\ref{eq-curl}) one 
obtains that vector $\vec{u}$ satisfies the following equation:

\begin{equation}
  \mathrm{curl} \; \vec{u} =
  \left[
    \vec{u} \, \times \,
    {\partial \vec{u} \over \partial t}
  \right]
\label{eq-anti-Euler}
\end{equation}
where $[ ... \times ...]$ is the usual three-dimensional wedge 
(vector) product.

Equation (\ref{eq-anti-Euler}), which can be termed as 'dual to' 
the Euler equation,
  $\partial \vec{u} / \partial t =
  \left[ \vec{u} \, \times \, \mathrm{curl} \; \vec{u} \right]$,
has already been discussed in the literature. For example, it has
been proposed as a relativistic generalization of the
  $\mathrm{curl} \, \vec{v}_{s} = 0$
condition for the velocity of a superfluid condensate (see, e.g., 
\cite{Putterman}). However, its relation with the integrable 
Volterra system (which also implies integrability of this model) 
seems to be new, and we are going to present more elaborated 
studies of (\ref{eq-anti-Euler}) in future papers.

\section{Soliton solutions.} \label{sec-solitons}

Here we would like to recall that equations are not only to be 
derived or classified, but should also be solved. In this section 
we present some class of solutions of the negative VH, namely the 
dark soliton ones.

There are many ways to derive pure-soliton solutions for an
integrable model: the inverse problem, the dressing method,
Backlund-Darboux tansforms, Hirota's ansatz. The key idea of the
last approach is that soliton solutions of all the integrable 
models possess the same structure, and the only thing one has to 
do while solving a particular equation is to determine some 
constants. In this paper we also exploit this fact. However, we do 
not use Hirota's technique but solve our equations using the 
elementary matrix calculus. Our starting point is that usually 
$N$-soliton solutions can be constructed of determinants of some 
combinations of the $N \times N$ matrices having the following 
structure:

\begin{equation}
  \left( A_{n} \right)_{jk} =
  { \ell_{j} a_{nk}(t) \over L_{j} - R_{k} }
\label{A-matrices}
\end{equation}
i.e. matrices which satisfy the equation

\begin{equation}
  L A_{n} - A_{n} R = | \,\ell\, \rangle \langle a_{n} |
\label{LAR}
\end{equation}
Here we use the 'bra-ket' notation,

\begin{equation}
  \langle a_{n} | = \left( a_{n,1}, ... , a_{n,N} \right)
\qquad
\qquad
  | \,\ell\, \rangle = \left( \ell_{1}, ... , \ell_{N} \right)^{T}
\end{equation}
and $L$ and $R$ are some diagonal matrices

\begin{equation}
  L = \mathrm{diag} \left( L_{1}, ... , L_{N} \right)
\qquad
\qquad
  R = \mathrm{diag} \left( R_{1}, ... , R_{N} \right)
\end{equation}

\subsection{Algebra of matrices (\ref{A-matrices}).}

Before proceeding further we present some formulae related to 
matrices (\ref{A-matrices}) which we need to construct soliton
solutions of the negative Volterra equations under the so-called
'finite-density' boundary conditions. We do not consider the most
general case of (\ref{LAR}), but restrict ourselves to

\begin{equation}
  L = R^{-1}
\end{equation}
and specify from the beginning the $n$-dependence by

\begin{equation}
  A_{n+1} = A_{n} R
\label{n-dependence}
\end{equation}
(as will be shown below, namely these matrices we need for our
purposes). From (\ref{LAR}), which we rewrite now as

\begin{equation}
  R^{-1} A_{n} - A_{n+1} = | \,\ell\, \rangle \langle a_{n} |
\label{LAR1}
\end{equation}
or

\begin{equation}
  A_{n} Z - Z A_{n+1} = Z | \,\ell\, \rangle \langle a_{n} | Z
\label{LAR2}
\end{equation}
where

\begin{equation}
  Z = \left( I + R^{-1} \right)^{-1}
\end{equation}
and $I$ is $N \times N$ unit matrix, one can derive that matrices
$F_{n}$ inverse to $I + A_{n}$,

\begin{equation}
  F_{n} = \left( I + A_{n} \right)^{-1}
\end{equation}
satisfy the identities

\begin{equation}
  F_{n+1}R^{-1} - R^{-1}F_{n-1} =
  F_{n+1} | \,\ell\, \rangle \langle a_{n-1} | F_{n-1}
\label{LFL}
\end{equation}
and

\begin{equation}
  Z F_{n+1} - F_{n} Z = Z F_{n+1} | \,\ell\, \rangle \langle a_{n} | F_{n} Z.
\label{ZFZ}
\end{equation}
These relations, together with the following formulae
\begin{eqnarray}
  1 - \langle a_{n} | F_{n} Z | \ell \rangle & = &
  { \omega_{n+1} \over \omega_{n} }
\label{alpha}
\\[1mm]
  1 + \langle a_{n-1} | Z F_{n}  | \ell \rangle  & = &
  { \omega_{n-1} \over \omega_{n} }
\label{beta}
\end{eqnarray}
where

\begin{equation}
  \omega_{n} = \mathrm{det} \left( I + A_{n} \right)
\label{def-omega}
\end{equation}
lead to
\begin{eqnarray}
  \omega_{n} \, F_{n} Z | \ell \rangle & = &
  \omega_{n+1} \, Z F_{n+1} | \ell \rangle
\label{f-ket}
\\
  \omega_{n} \, \langle a_{n} | F_{n} Z & = &
  \omega_{n+1} \, \langle a_{n} | Z F_{n+1}.
\label{f-bra}
\end{eqnarray}
Formulae (\ref{alpha}) and (\ref{beta}) follow from
(\ref{n-dependence}) and the fact that for any bra-vector (i.e.
$N$-row) $\langle u |$ and any ket-vector (i.e. $N$-column) $| v
\rangle$,
  $ \mathrm{det} \left( I + | v \rangle \langle u | \, \right) =
  1 + \langle u | v \rangle $,
while (\ref{f-ket}) and (\ref{f-bra}) can be obtained from
(\ref{ZFZ}) by multiplying it by
  $| \ell \rangle$ and $\langle a_{n} |$.
Formulae (\ref{LFL})--(\ref{f-bra}) are all we need in the 
following consideration.

\subsection{Auxiliary sytem.}

Now we are ready to solve auxiliary equations

\begin{equation}
  \bar\partial_{1} \ln { \omega_{n+1} \over \omega_{n-1} } =
  { \omega_{n}^{2} \over \omega_{n+1}\omega_{n-1} }  - 1
\label{aux-first}
\end{equation}
and

\begin{equation}
  \bar\partial_{j+1} \ln { \omega_{n+1} \over \omega_{n-1} } =
  - { \omega_{n}^{2} \over \omega_{n+1}\omega_{n-1} }
  \bar\partial_{1j} \ln \omega_{n}.
\label{aux-hie}
\end{equation}
Solutions of these equations then can be easily modified to become
solutions of the negative VH under the non-vanishing boundary
conditions $u_{n} \to u_{\infty}$ as $n \to \pm\infty$

The crucial point of the ansatz (\ref{A-matrices}) is that
dependence on times $\bar{t}_{1}$, $\bar{t}_{2}$, ... is
incorporated in the $N$-rows
  $\langle a_{n} | =
   \langle a_{n} \left(\bar{t}_{1}, \bar{t}_{2}, ... \right) | $.
Moreover, we assume (which is usual for pure soliton solutions) 
that $\ln a_{nk}$ is a {\it linear} function of times. Hence, 
differentiating the $A$-matrices leads to multiplication from the 
right by some constant matrices, $C_{j}$,

\begin{equation}
  {\partial \over \partial \bar{t}_{j}} \; A_{n}  =  A_{n} C_{j}
\label{dA}
\end{equation}
which should be determined. As follows from (\ref{dA}),

\begin{equation}
  {\partial \over \partial \bar{t}_{j}} \; \ln\omega_{n} =
  \mathrm{tr} \; F_{n} A_{n} C_{j} =
  \mathrm{tr} \; \left( I - F_{n} \right) C_{j}
\label{domega}
\end{equation}
which gives

\begin{equation}
  {\partial \over \partial \bar{t}_{j}} \;
  \ln { \omega_{n+1} \over \omega_{n-1} } =
  \mathrm{tr} \; \left( F_{n-1} - F_{n+1} \right) C_{j}.
\end{equation}
Using (\ref{LFL}) and the fact that $R$ and $C_{j}$ commute, the
last equation can be rewritten as
\begin{eqnarray}
  {\partial \over \partial \bar{t}_{j}} \;
  \ln { \omega_{n+1} \over \omega_{n-1} }
  & = &
  \mathrm{tr} \; \left( R^{-1}F_{n-1} - F_{n+1}R^{-1} \right) C_{j}R
\\ & = &
  - \langle a_{n-1} | F_{n-1} C_{j}R F_{n+1} | \ell \rangle
\label{domega-ratio}
\end{eqnarray}
On the other hand, by multiplying (\ref{alpha}) and (\ref{beta}) 
one can derive the identity

\begin{equation}
  { \omega_{n}^{2} \over \omega_{n-1} \omega_{n+1} } =
  1 -
  \langle a_{n-1} | F_{n-1} X_{n} F_{n+1} | \ell \rangle
\end{equation}
where

\begin{equation}
  X_{n} =
  Z \left( I + A_{n+1} \right) -
  \left( I + A_{n-1} \right) RZ +
  Z | \ell \rangle  \langle a_{n} | Z
\end{equation}
Applying (\ref{LFL}) one gets

\begin{equation}
  X_{n} =
  \left( I - R \right) Z = \mathrm{constant}
\end{equation}
i.e.

\begin{equation}
  { \omega_{n}^{2} \over \omega_{n-1} \omega_{n+1} } - 1 =
  \langle a_{n-1} | F_{n-1} (R-I)Z F_{n+1} | \ell \rangle
\end{equation}
Comparing (\ref{domega-ratio}) and the last formula one can
conclude that $\omega_{n}$ solve (\ref{aux-first}) if

\begin{equation}
  C_{1} = \left( R^{-1} - I \right) Z
\end{equation}
or

\begin{equation}
  C_{1} = (I - R)(I + R)^{-1}.
\label{C1}
\end{equation}

Now our aim is to solve (\ref{aux-hie}). Differentiating
(\ref{domega}) with respect to $\bar{t}_{1}$ we get

\begin{equation}
  {\partial^{2} \over \partial \bar{t}_{1} \partial \bar{t}_{j}} \;
  \ln\omega_{n}  =
  \mathrm{tr} \; A_{n} F_{n} C_{1} F_{n} C_{j}.
\label{dif2omega}
\end{equation}
From definition (\ref{C1}) of $C_{1}$ and (\ref{LAR2}) one can
derive

\begin{equation}
  A_{n} C_{1} =
  Z A_{n} - A_{n} Z +
  Z | \ell \rangle \langle a_{n-1} | Z
\end{equation}
which leads to

\begin{equation}
  A_{n} F_{n} C_{1} F_{n} =
  F_{n} Z - Z F_{n} +
  F_{n} Z | \ell \rangle \langle a_{n-1} | Z F_{n}.
\end{equation}
Noting that $Z$ and $C_{j}$ commute and that the trace of a
commutator is zero, the right-hand side of (\ref{dif2omega}) can
be rewritten as

\begin{equation}
  \mathrm{tr} \; A_{n} F_{n} C_{1} F_{n} C_{j} =
  \langle a_{n-1} | Z F_{n} C_{j} F_{n} Z | \ell \rangle
\end{equation}
which gives, together with (\ref{f-ket}) and (\ref{f-bra}),

\begin{equation}
  {\partial^{2} \over \partial \bar{t}_{1} \partial \bar{t}_{j}} \;
  \ln\omega_{n}  =
  \frac{\omega_{n-1}\omega_{n+1}}{\omega_{n}^{2}}
  \langle a_{n-1} | F_{n-1} C_{j} Z^{2} F_{n+1} | \ell \rangle.
\end{equation}
Comparing this expression and equation (\ref{domega-ratio}) for
the derivative with respect to the ($j+1$)th time, 
$\bar{t}_{j+1}$, one can conclude that $\omega_{n}$ is a solution 
of (\ref{aux-hie}), if matrices $C_{j+1}$ and $C_{j}$ are related 
by

\begin{equation}
  C_{j+1} = C_{j} Z^{2}R^{-1}.
\end{equation}
This recurence together with the 'initial condition' (\ref{C1}) 
can be easily solved:

\begin{equation}
  C_{j} =
  C_{1} \left( R + R^{-1} + 2I \right)^{1-j} =
  C_{1} \left( R^{1/2} + R^{-1/2} \right)^{2-2j}.
\label{Cj}
\end{equation}

In such a way we have shown that if the time-dependence of the 
$A$-matrices is given by (\ref{dA}) and (\ref{Cj}), then the 
quantities $\omega_{n}$ given by (\ref{def-omega}) solve 
(\ref{aux-hie}).

\subsection{Dark-soliton solutions of the VH.}

It is easy to see that equations (\ref{aux-first}) and
(\ref{aux-hie}) differ only in a constant term on the right-hand
side of (\ref{aux-first}) , which can be taken into account by
$\exp\left(n \bar{t}_{1}/2\right)$ multiplier: functions

\begin{equation}
  \tau_{n} = \exp\left({ n \bar t_{1} \over 2 } \right)\omega_{n}
\end{equation}
solve (\ref{neg-hie-1}). The limiting values of the coresponding
$u$-functions are unity and to meet general 'finite-density' 
boundary conditions,

\begin{equation}
  \lim_{n \to \pm\infty} u_{n} = u_{\infty}
\end{equation}
one has to add the $u_{\infty}^{n^{2}/4}$ factor and to rescale
the times, $\bar{t}_{j} \to \bar{t}_{j} / u_{\infty}^{j-1/2}$. The 
final formulae for the $N$-soliton solutions of the VH can be 
written as

\begin{equation}
  \tau_{n} =
  u_{\infty}^{n^{2}/4}
  \exp\left( { n \bar t_{1} \over 2\sqrt{u_{\infty}} } \right)
  \det\left|
    1 +
    { R_{k}^{n} \, a_{k}(\bar t) \over R_{j}^{-1} - R_{k} }
  \right|_{j,k=1,...,N}.
\end{equation}
The time dependence of $a_{k}$ on negative times is given by

\begin{equation}
  a_{k}(\bar t) = a_{k}\left( \bar t_{1}, \bar t_{2}, ... \right) =
  a_{k}^{(0)} \exp\left(
    \sum_{j=1}^{\infty} \nu_{k}^{(j)} \, \bar t_{j}
  \right)
\end{equation}
with

\begin{equation}
  \nu_{k}^{(j)} =
  { 1 \over \sqrt{u_{\infty}} }
  { 1 - R_{k} \over 1 + R_{k} }
  \left[ u_{\infty} \left( 2 + R_{k} + R_{k}^{-1} \right)\right]^{1-j}
\end{equation}
where $R_{k}$, $a_{k}^{(0)}$, $k=1,...,N$ are arbitrary constants.
Note that constants $\ell_{j}$ appearing in (\ref{A-matrices})
have been incorporated in $a_{k}^{(0)}$ by the transform
  $A_{n} \to M^{-1} A_{n} M$
with
  $M = \mathrm{diag} \left(\ell_{1}, ..., \ell_{N}\right)$
which does not change determinants.

\section{Conclusion.}

In the present paper we studied the negative Volterra flows. The 
main result of this work is that even in the theory of, so to say, 
classical systems there are some questions which have not been 
discussed in the literature yet. We applied the widely used zero 
curvature approach to the well-known scattering problem and 
obtained some results which seem to be new. Deriving equations 
(\ref{neg-hie}) and (\ref{neg-hie-1}) or (\ref{hie-Hirota}) and 
(\ref{hie-Hirota-1}) is not a very difficult problem, though it 
fills some gaps in the theory of one of the oldest integrable 
systems. However, a hierarchy is more than a set of equations, and 
one can hardly enumerate all equations which are contained in this 
or that hierarchy, i.e. to study all possible differential 
consequences of equations of a given hierarchy. In section 
\ref{sec-models} we demonstrated that from the extended VH, which 
is a set of differential-difference equations, some of which are 
nonlocal, one can extract rather different systems. For example, 
in our opinion it is very difficult to say \textit{a priori} that 
equation (\ref{eq-anti-Euler}) (a vector partial differential 
equation in 3+1 dimensions) has any relation to the Volterra 
chain. In future papers we are going to present some other 
integrable systems which can be 'embedded' into (or extracted 
from) the extended VH, and we hope that the results presented 
above demonstrate that the theory of integrable systems is far 
from finished.

\section*{Acknowledgements}

VEV is supported by Ministerio de Educaci\'on, Cultura y Deporte 
under grant SAB2000-0256.

\section*{Appendix.}

The question of compatibility of the equations of a hierarchy (or, 
in other words, of commutativity of the corresponding flows) is of 
vital importance and usually not trivial. The most straightforward 
approach to this problem is to derive the Hamiltonians generating 
these flows and to show that they are in involution. However, the 
Hamiltonian representation of the negative Volterra hierarchy is 
still unknown and is a subject of future studies. Thus here we do 
not discuss this rather serious problem in the most general way 
and restrict ourselves to proving the commutativity of the first 
(classical) Volterra flow (\ref{volterra-chain}) and all the 
negative flows derived in this paper, a fact which is crucial for 
what has been presented in section \ref{sec-models}.

This claim can be verified directly in the simlest cases. Let us 
first consider the Volterra equation (\ref{volterra-chain}), 
rewriting it now as

\begin{equation}
  { \partial \over \partial t_{1} } \;
  \ln \frac{\tau_{n}}{\tau_{n-1}}  = u_{n},
\label{app-pos-1}
\end{equation}
which in its turn leads to 

\begin{equation}
  { \partial \over \partial t_{1} } \;
  \ln \frac{\tau_{n+1}}{\tau_{n-1}}  = u_{n+1} + u_{n},
\end{equation}
and show that it is compatible with (\ref{neg-hie-1}),

\begin{equation}
  { \partial \over \partial\bar t_{1} } \;
  \ln \frac{\tau_{n+1}}{\tau_{n-1}}  = \frac{1}{p_{n}}.
\label{app-neg-1}
\end{equation}
To do this we need auxiliary formulae for the derivatives of 
$u_{n}$ and $p_{n}$. The first one (which follows from 
(\ref{app-neg-1})) is

\begin{equation}
  { \partial \over \partial\bar t_{1} } \; \ln u_{n} =
  { \partial \over \partial\bar t_{1} } \; 
  \left(
    \ln \frac{\tau_{n+1}}{\tau_{n-1}} -
    \ln \frac{\tau_{n}}{\tau_{n-2}} 
  \right) =
  \frac{1}{p_{n}} - \frac{1}{p_{n-1}}
\label{app-d-u}
\end{equation}
while the second one is a consequence of (\ref{app-pos-1}):

\begin{equation}
  { \partial \over \partial t_{1} } \; \ln p_{n} =
  { \partial \over \partial t_{1} } \; 
  \left(
    \ln \frac{\tau_{n+1}}{\tau_{n}} -
    \ln \frac{\tau_{n}}{\tau_{n-1}} 
  \right) =
  u_{n+1} - u_{n}.
\label{app-dbar-p}
\end{equation}
Now it is easy to derive

\begin{equation}
  { \partial \over \partial\bar t_{1} } 
  \left(
    { \partial \over \partial t_{1} } \; 
    \ln \frac{\tau_{n+1}}{\tau_{n-1}} 
  \right)
  =
  u_{n+1} \left( \frac{1}{p_{n+1}} - \frac{1}{p_{n}}   \right) +
  u_{n}   \left( \frac{1}{p_{n}}   - \frac{1}{p_{n-1}} \right)
  =
  - p_{n+1} + p_{n-1}
\label{app-dbard-1}
\end{equation}
(we have used (\ref{app-d-u}) and the fact that 
  $u_{n} = p_{n} p_{n-1}$).

On the other hand, it follows from (\ref{app-neg-1}) and 
(\ref{app-dbar-p})

\begin{equation}
  { \partial \over \partial t_{1} } 
  \left(
    { \partial \over \partial\bar t_{1} } \; 
    \ln \frac{\tau_{n+1}}{\tau_{n-1}} 
  \right)
  =
  \frac{1}{p_{n}} \left( u_{n} - u_{n+1} \right) 
  =
  p_{n-1} - p_{n+1}.
\label{app-ddbar-1}
\end{equation}
Since the right-hand sides of (\ref{app-dbard-1}) and 
(\ref{app-ddbar-1}) are equal, one can conclude that 

\begin{equation}
  { \partial \over \partial\bar t_{1} } \; 
  { \partial \over \partial t_{1} } \;
  \ln \frac{\tau_{n+1}}{\tau_{n-1}} 
  =
  { \partial \over \partial t_{1} } \;
  { \partial \over \partial\bar t_{1} } \; 
  \ln \frac{\tau_{n+1}}{\tau_{n-1}} 
\end{equation}
which proves the compatibility of (\ref{app-pos-1}) and 
(\ref{app-neg-1}).

Further we proceed by induction. Suppose that Volterra flow 
(\ref{app-pos-1}) commutes with all the negative flows
  $\partial / \partial\bar t_{1}$,
  $\partial / \partial\bar t_{2}$, ... ,
  $\partial / \partial\bar t_{j}$.
Now our aim is to show that it commutes with the $(j+1)$th 
negative flow as well. Indeed, differentiating equation 
(\ref{neg-hie}), which can be rewritten as

\begin{equation}
  { \partial \over \partial \bar t_{j+1} } \,
  \ln { \tau_{n+1} \over \tau_{n-1} } 
  =
  - \frac{1}{p_{n}} \,
  {\partial^{2} \over \partial\bar t_{1} \partial\bar t_{j} }
  \ln \tau_{n} 
\label{app-dbar-jone}
\end{equation}
with respect to $t_{1}$, one can obtain

\begin{eqnarray}
  { \partial \over \partial t_{1} } 
  \left(
    { \partial \over \partial \bar t_{j+1} } \,
    \ln { \tau_{n+1} \over \tau_{n-1} } 
  \right)
  & = &
  - { \partial \over \partial t_{1} } 
  \left( 
    \frac{1}{p_{n}} \,
    {\partial^{2} \over \partial\bar t_{1} \partial\bar t_{j} }
    \ln \tau_{n} 
  \right)
\cr
  & = &
  - \left( { \partial \over \partial t_{1} } \, \frac{1}{p_{n}} \right)
    {\partial^{2} \over \partial\bar t_{1} \partial\bar t_{j} }
    \ln \tau_{n} 
  - \frac{1}{p_{n}} 
    { \partial \over \partial \bar t_{j} } 
    \left( 
      {\partial^{2} \over \partial t_{1} \partial\bar t_{1} }
      \ln \tau_{n} 
    \right)
\cr
  & = & 
  \left( p_{n+1} - p_{n-1} \right)
  {\partial^{2} \over \partial\bar t_{1} \partial\bar t_{j} }
    \ln \tau_{n} 
  + 
  \frac{1}{p_{n}} 
  { \partial \over \partial \bar t_{j} } \, p_{n}
\label{app-ddbar-j}
\end{eqnarray}
Here we have used the commutativity of 
  $ \partial / \partial t_{1} $ and $ \partial / \partial t_{j} $
as well as the identity

\begin{equation}
  {\partial^{2} \over \partial t_{1} \partial\bar t_{1} }
    \ln \tau_{n} 
  = - p_{n}
\end{equation}
which can be derived by differentiating (\ref{app-neg-1}) with 
respect to $t_{1}$ (see section \ref{model-Toda}). Noting that

\begin{equation}
  { \partial \over \partial\bar t_{j} } \, p_{n} =
  - p_{n}^{2} \;
    { \partial \over \partial\bar t_{j} } \, \frac{1}{p_{n}} =
  - p_{n}^{2} \;  
    {\partial^{2} \over \partial\bar t_{1} \partial\bar t_{j} } \;
    \ln \frac{\tau_{n+1}}{\tau_{n-1}} 
\end{equation}
one can rewrite (\ref{app-ddbar-j}) as

\begin{equation}
  { \partial \over \partial t_{1} } 
  \left(
    { \partial \over \partial \bar t_{j+1} } \,
    \ln { \tau_{n+1} \over \tau_{n-1} } 
  \right)
  = 
  \left( p_{n+1} - p_{n-1} \right)
    {\partial^{2} \over \partial\bar t_{1} \partial\bar t_{j} } \,
    \ln \tau_{n} 
  +
  p_{n} 
  {\partial^{2} \over \partial\bar t_{1} \partial\bar t_{j} } \,
    \ln \frac{\tau_{n-1}}{\tau_{n+1}}. 
\label{app-ddbar-jone}
\end{equation}
On the other hand,

\begin{equation}
  { \partial \over \partial\bar t_{j+1} } 
  \left(
    { \partial \over \partial t_{1} } \,
    \ln { \tau_{n+1} \over \tau_{n-1} } 
  \right)
  = 
  { \partial \over \partial\bar t_{j+1} } 
  \left( u_{n+1} + u_{n} \right).
\label{app-ddbar-ju}
\end{equation}
One can obtain from (\ref{app-dbar-jone}) the following formula 
for the derivative of $u_{n}$:

\begin{eqnarray}
  { \partial \over \partial\bar t_{j+1} } \; u_{n} & = &
  p_{n} p_{n-1} { \partial \over \partial\bar t_{j+1} } 
  \left( 
    \ln { \tau_{n+1} \over \tau_{n-1} } -
    \ln { \tau_{n} \over \tau_{n-2} }
  \right)
\cr 
  & = & 
  p_{n} 
  {\partial^{2} \over \partial\bar t_{1} \partial\bar t_{j} } \,
    \ln \tau_{n-1} 
  -
  p_{n-1} 
  {\partial^{2} \over \partial\bar t_{1} \partial\bar t_{j} } \,
    \ln \tau_{n} 
\end{eqnarray}
Substituting this expression in (\ref{app-ddbar-ju}) leads to

\begin{equation}
  { \partial \over \partial\bar t_{j+1} } 
  \left(
    { \partial \over \partial t_{1} } \,
    \ln { \tau_{n+1} \over \tau_{n-1} } 
  \right)
  = 
  \left( p_{n+1} - p_{n-1} \right)
  {\partial^{2} \over \partial\bar t_{1} \partial\bar t_{j} } \,
    \ln \tau_{n} 
  +
  p_{n} 
  {\partial^{2} \over \partial\bar t_{1} \partial\bar t_{j} } \,
    \ln \frac{\tau_{n-1}}{\tau_{n+1}}  
\label{app-dbard-jone}
\end{equation}
Comparing the right-hand sides of (\ref{app-ddbar-jone}) and 
(\ref{app-dbard-jone}) one can easily see that they coincide, 
which means that

\begin{equation}
  { \partial \over \partial t_{1} } \,
  { \partial \over \partial\bar t_{j+1} } \;
  \ln { \tau_{n+1} \over \tau_{n-1} } 
  = 
  { \partial \over \partial\bar t_{j+1} } \,
  { \partial \over \partial t_{1} } \;
  \ln { \tau_{n+1} \over \tau_{n-1} } 
\end{equation}
This leads by induction to the fact that all the negative Volterra 
flows $\partial / \partial\bar t_{j} $ commute with the classical 
one, $\partial / \partial t_{1} $.


\end{document}